\documentclass[iop,numberedappendix]{emulateapj}
\usepackage{natbib}
\usepackage{graphicx}
\usepackage{rotating}
\usepackage{verbatim}
\usepackage{amssymb}

\slugcomment{Revised version September 19 2011 -- Accepted for publication in ApJL}

\shorttitle{Black hole versus total stellar mass scaling relations}
\shortauthors{CISTERNAS ET AL.}

\begin{document}

\title{Secular Evolution and a Non-evolving Black Hole to Galaxy Mass Ratio\\ in the Last 7 G\lowercase{yr}$^{\star}$}


\author{Mauricio Cisternas$^{1,}$\altaffilmark{7},
 Knud Jahnke$^{1}$,
 Angela Bongiorno$^{2}$,
 Katherine J. Inskip$^{1}$,
 Chris D. Impey$^{3}$,\\
 Anton M. Koekemoer$^{4}$,
 Andrea Merloni$^{2}$,
 Mara Salvato$^{5}$,
 and Jonathan R. Trump$^{6}$
}
\email{cisternas@mpia.de}


\affil{$^{1}$ Max-Planck-Institut f\"{u}r Astronomie, K\"{o}nigstuhl 17, D-69117 Heidelberg, Germany}
\affil{$^{2}$ Max-Planck-Institut f\"{u}r Extraterrestrische Physik, Giessenbachstrasse 1, D-85748 Garching bei M\"{u}nchen, Germany}
\affil{$^{3}$ Steward Observatory, University of Arizona, 933 North Cherry Avenue, Tucson, AZ 85721, USA}
\affil{$^{4}$ Space Telescope Science Institute, 3700 San Martin Drive, Baltimore, MD 21218, USA}
\affil{$^{5}$ Max-Planck-Institut f\"{u}r Plasmaphysik, Boltzmanstrasse 2, D-85741 Garching, Germany}
\affil{$^{6}$ University of California Observatories/Lick Observatory, University of California, Santa Cruz, CA 95064, USA}

\altaffiltext{$\star$}{Based on observations with the NASA/ESA {\em Hubble Space Telescope}, obtained at the Space Telescope Science Institute, which is operated by AURA Inc, under NASA contract NAS 5-26555;
the XMM-Newton, an ESA science mission with instruments and contributions directly funded by ESA Member States and NASA;
and on data collected at the Magellan Telescope, which is operated by the Carnegie Observatories.
}

\altaffiltext{7}{Member of the International Max Planck Research School for Astronomy \& Cosmic Physics at the University of Heidelberg (IMPRS-HD), Germany}


\begin{abstract}
We present new constraints on the ratio of black hole (BH) mass to total galaxy stellar mass at $0.3<z<0.9$ for a sample of 32 type-1 active galactic nuclei (AGNs) from the XMM-COSMOS survey covering the range $M_{\mathrm{BH}}\sim10^{7.2-8.7}\,M_{\odot}$.
Virial $M_{\mathrm{BH}}$ estimates based on H$\beta$ are available from the COSMOS Magellan/IMACS survey.
We use high-resolution {\it Hubble Space Telescope (HST)} imaging to decompose the light of each type-1 AGN and host galaxy, and employ a specially-built mass-to-light ratio to estimate the stellar masses ($M_{\ast}$).
The $M_{\mathrm{BH}}-M_{\ast}$ ratio shows a zero offset with respect to the local relation for galactic bulge masses, and we also find no evolution in the mass ratio $M_{\mathrm{BH}}/M_{\ast}\varpropto(1+z)^{0.02\pm0.34}$ up to $z\sim0.9$.
Interestingly, at the high-$M_{\mathrm{BH}}$ end there is a positive offset from the $z=0$ relation, which can be fully explained by a mass function bias with a cosmic scatter of $\sigma_{\mu}=0.3$, reaffirming that the intrinsic distribution is consistent with zero evolution.
From our results we conclude that since $z\sim0.9$ no substantial addition of stellar mass is required:
the decline in star formation rates and merger activity at $z<1$ support this scenario.
Nevertheless, given that a significant fraction of these galaxies show a disk component, their bulges are indeed undermassive.
This is a direct indication that for the last 7 Gyr the only essential mechanism required in order that these galaxies obey the $z=0$ relation is a redistribution of stellar mass to the bulge, likely driven by secular processes, i.e., internal instabilities and minor merging.
\end{abstract}



\section{Introduction}

Observations of nearby galaxies have revealed the existence of a tight correlation between the masses of galactic bulges and their central supermassive black holes \citep[][, hereafter HR04]{mbh_m1, mbh_m2}.
While this correlation can be accounted for by a statistical convergence of typically several mergers per galaxy over cosmic time \citep{peng07, jahnke&maccio11}, the observed coupling has often been taken as an indication of physically driven co-evolution \citep{volonteri03}.

A strong constraint for either globally or individually coupled growth of BH and stellar mass is the evolution of their scaling relations with redshift.
Different theoretical models predict different levels of evolution \citep[e.g.,][]{granato04,robertson06,dimatteo08a}, and observations which directly probe the physical mechanisms that regulate this co-evolution are scarce.

A handful of studies probing the scaling relations beyond the local universe have found a larger ratio of BH mass to bulge stellar mass \citep[e.g.,][]{walter04,peng06a,treu07,jahnke09,merloni10,bennert10,decarli10}, suggesting that BHs grow earlier than their host spheroids.
Nevertheless, small number statistics and frequently ignored selection biases remain the main obstacles against additional evidence and more solid constraints.

In this letter we explore the $M_{\mathrm{BH}}-M_{\ast}$ relation for 32 type-1 AGNs in the redshift range $0.3<z<0.9$, with virial $M_{\mathrm{BH}}$ measurements.
Stellar masses are computed by combining the host galaxy luminosities (accurately measured via high-resolution {\it HST}/ACS imaging) with a specially built mass-to-light ratio ($M_{\ast}/L$) based on a large sample of type-2 AGNs, for which the stellar masses have been estimated through a spectral energy distribution (SED) fitting.

Throughout we assume a flat cosmology with $H_0=70\mathrm{km\,s}^{-1}\mathrm{Mpc}^{-1}$, $\Omega_{M}=0.3$, and $\Omega_{\Lambda}=0.7$.

\begin{deluxetable*}{rcccccc}
\tabletypesize{\scriptsize}
\tablecaption{Type-1 AGN Sample and Derived Host Galaxy Properties.\label{tbl-1}}
\tablewidth{1\textwidth}
\tablehead{
Object (J2000)\hspace{1cm} & $z$ & log $M_{\mathrm{BH}}$$^{\mathrm{a}}$ & S\'{e}rsic $n^{\mathrm{b}}$ & $F814W_{\mathrm{Host}}$ & log $L_{F814W}$$^{\mathrm{c}}$ & log $M_{\ast}$$^{\mathrm{d}}$\\
  &   & ($M_{\odot}$) &   & (AB) & ($L_{\odot}$) & ($M_{\odot}$)
}
\startdata
COSMOS J095817.54+021938.5 &    0.73 &    7.72 &   1.0 &   21.82 &    8.98 &   10.30 \\
SDSS J095819.88+022903.6 &    0.34 &    8.29 &   1.4 &   17.51 &   10.03 &   11.23 \\
COSMOS J095831.65+024901.6 &    0.34 &    8.08 &   1.4 &   19.33 &    9.29 &   10.65 \\
COSMOS J095840.61+020426.6 &    0.34 &    8.39 &   1.8 &   18.16 &    9.76 &   11.02 \\
COSMOS J095845.80+024634.0 &    0.35 &    7.39 &   2.8 &   19.70 &    9.16 &   10.54 \\
SDSS J095902.76+021906.5 &    0.34 &    8.66 &   4.0 &   17.81 &    9.91 &   11.14 \\
COSMOS J095909.53+021916.5 &    0.38 &    7.77 &   2.0 &   19.44 &    9.34 &   10.68 \\
COSMOS J095928.31+022106.9 &    0.35 &    7.24 &   1.0 &   18.41 &    9.68 &   10.95 \\
COSMOS J100002.21+021631.8 &    0.85 &    8.29 &   4.0 &   19.63 &    9.99 &   11.07 \\
SDSS J100012.91+023522.8 &    0.70 &    8.15 &   4.0 &   19.01 &   10.07 &   11.17 \\
COSMOS J100014.55+023852.7 &    0.44 &    7.79 &   4.0 &   20.05 &    9.23 &   10.57 \\
COSMOS J100017.54+020012.6 &    0.35 &    7.59 &   2.3 &   19.98 &    9.07 &   10.47 \\
SDSS J100025.25+015852.2 &    0.37 &    8.58 &   4.0 &   19.75 &    9.21 &   10.57 \\
COSMOS J100028.63+025112.7 &    0.77 &    8.49 &   4.0 &   20.13 &    9.70 &   10.86 \\
COSMOS J100029.69+022129.7 &    0.73 &    8.03 &   1.0 &   19.58 &    9.87 &   11.01 \\
COSMOS J100033.38+015237.2 &    0.83 &    8.07 &   1.1 &   20.42 &    9.65 &   10.81 \\
COSMOS J100033.49+013811.6 &    0.52 &    8.01 &   4.0 &   20.49 &    9.21 &   10.54 \\
COSMOS J100037.29+024950.6 &    0.73 &    7.41 &   1.0 &   21.65 &    9.05 &   10.36 \\
SDSS J100043.15+020637.2 &    0.36 &    8.07 &   4.0 &   17.44 &   10.10 &   11.28 \\
COSMOS J100046.72+020404.5 &    0.55 &    7.75 &   0.7 &   18.89 &    9.91 &   11.08 \\
COSMOS J100058.71+022556.2 &    0.69 &    7.91 &   1.0 &   20.61 &    9.42 &   10.66 \\
COSMOS J100118.52+015543.0 &    0.53 &    8.22 &   1.0 &   19.58 &    9.59 &   10.84 \\
COSMOS J100141.09+021300.0 &    0.62 &    7.35 &   3.4 &   20.82 &    9.24 &   10.53 \\
COSMOS J100146.49+020256.7 &    0.67 &    7.73 &   4.0 &   20.27 &    9.52 &   10.75 \\
COSMOS J100202.22+024157.8 &    0.79 &    8.24 &   1.0 &   21.24 &    9.29 &   10.53 \\
COSMOS J100205.03+023731.5 &    0.52 &    8.38 &   4.0 &   18.54 &    9.99 &   11.15 \\
COSMOS J100212.11+014232.4 &    0.37 &    7.70 &   1.8 &   20.02 &    9.09 &   10.48 \\
COSMOS J100218.32+021053.1 &    0.55 &    8.61 &   1.9 &   18.50 &   10.06 &   11.20 \\
COSMOS J100230.06+014810.4 &    0.63 &    7.50 &   1.0 &   20.20 &    9.49 &   10.73 \\
COSMOS J100230.65+024427.6 &    0.82 &    7.82 &   2.3 &   20.54 &    9.59 &   10.77 \\
SDSS J100232.13+023537.3 &    0.66 &    8.19 &   1.8 &   19.36 &    9.87 &   11.03 \\
COSMOS J100243.96+023428.6 &    0.38 &    8.25 &   4.0 &   18.15 &    9.86 &   11.08
\enddata
\tablenotetext{a}{Uncertainty quoted as 0.4 dex \citep{cosmos_trump09b}.}
\tablenotetext{b}{Selected from the fixed ($n=$1,4) and free fits.}
\tablenotetext{c}{As defined in Equation~(\ref{lum}).}
\tablenotetext{d}{Total propagated uncertainty is 0.35 dex.}
\end{deluxetable*}

\section{Sample}

Our type-1 AGN sample is based on the XMM-COSMOS survey catalog \citep{cosmos_xmm1, cosmos_xmm2}, which consists of $\sim$1800 bright X-ray sources.
Accurate identification of their optical counterparts and multiwavelength properties are described in detail in \citet{cosmos_brusa10}.
Targeted spectroscopic observations with Magellan/IMACS \citep{cosmos_trump09a} and publicly available spectra from SDSS \citep{schneider07} allowed the accurate classification of $\sim400$ intermediate-luminosity type-1 AGNs.
Within the redshift range $0.3<z<0.9$, we selected those with available virial $M_{\mathrm{BH}}$ estimates from \citet{cosmos_trump09b}, based on measurements of the width of the H$\beta$ line applied on the scaling relations of \citet{vestergaard06}.
This selection results in 32 type-1 AGNs.

A fundamental part of our analysis of the AGN host galaxies is based on the {\it HST}/ACS imaging of the COSMOS field \citep{cosmos_hst,cosmos_acs};
this is therefore an ideal redshift range.
Beyond $z\sim1$, the F814W filter shifts to restframe UV, where the point-like AGN starts to dominate the overall light distribution, making it highly difficult to resolve its host galaxy.

In Table~\ref{tbl-1} we summarize the sample, including the properties derived in the following sections.

\begin{figure*}[t]
\centering
\resizebox{0.9\textwidth}{!}{\includegraphics{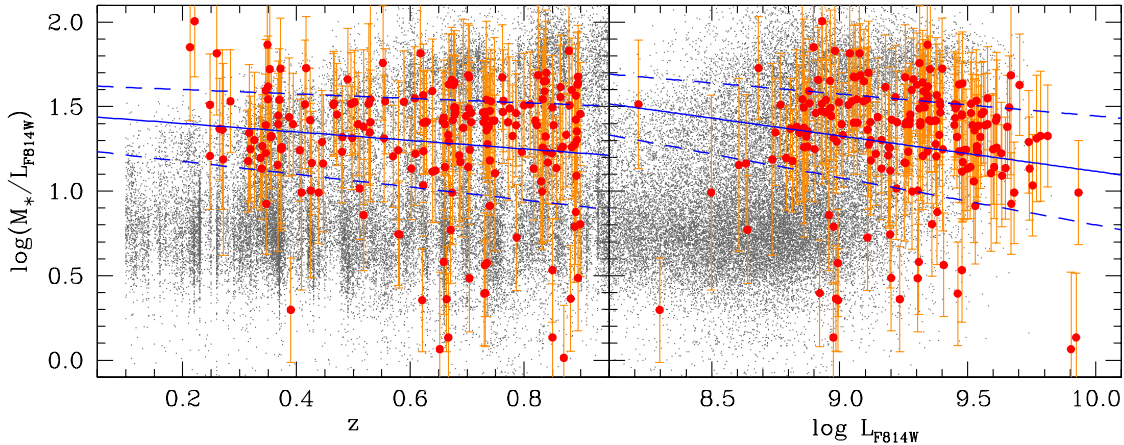}}
\caption{Mass-to-light as a function of redshift (left panel) and luminosity (right panel) for the type-2 AGN sample, plotted as the filled circles.
The dots correspond to $\sim$40,000 inactive galaxies with $I<23$ from the COSMOS catalog \citep{cosmos_ilbert09}, shown for comparison.
The solid lines are the projections of the best-fit plane from Equation~(\ref{m2l}) at the median luminosity and redshift values respectively.
The typical $1\sigma$ errors on the fit are marked by the dashed lines.\label{fig_m2l}}
\end{figure*}

\section{AGN Host Galaxy Masses}
We infer the stellar masses for our 32 type-1 AGN host galaxies under the generally accepted unified AGN model \citep{unification1}, in which type-1 and type-2 sources correspond to different viewing angles of the same phenomena.
While recent evidence shows that AGN type depends on accretion rate as well as orientation \citep{trump11}, at $z\sim1$ and moderate nuclear luminosities the host galaxies of both types of AGNs show equivalent properties \citep{ammons11}.
We therefore use a $M_{\ast}/L$ especially built from a large sample of type-2 AGN host galaxies with stellar masses derived from SED fitting.
To quantify the light of the type-1 AGN hosts we model and remove the flux contribution of the active nuclei, utilizing high-resolution {\it HST}/ACS imaging with the F814W filter and 0\farcs03/pixel sampling.
The motivation for this approach is the simplicity and strength of the technique, which might prove extremely useful for cases in which the spectral coverage on the host galaxy is limited or even non-existent.

\subsection{Assessing a characteristic AGN mass-to-light ratio}

With the goal of estimating stellar masses for our type-1 AGN host galaxies, we need to construct realistic $M_{\ast}/L$ distributions.
AGN hosts have characteristically young stellar populations different from the general population \citep{kauffmann03,sanchez04,jahnke04a,jahnke04b}, and are drawn from the bright-end of the galaxy luminosity function \citep{zakamska06}.

With this in mind, we build a $M_{\ast}/L$ from a large sample of type-2 AGNs, also from the COSMOS survey.
Type-2 AGNs have the advantage that the bright nucleus is highly obscured, leaving the host galaxy almost free of contamination from AGN light.
This allows us to exploit the vast ground-based multiwavelength photometry from COSMOS to accurately model the host galaxy SED, providing well-determined stellar masses.

Our sample of spectroscopically confirmed type-2 AGNs \citep{cosmos_trump09a,cosmos_bongiorno10} consists of 199 sources at $z\sim0.2-0.9$.
While the AGN contribution to the overall SED is minimal, for the sake of accuracy a large grid of composite galaxy+AGN models were used to find the best fit template most fully representing the type-2 system (Bongiorno et al. in prep).
Stellar masses were derived from the SEDs, assuming a Chabrier initial mass function (IMF).
Comparison of these stellar masses with fits of just a single galaxy template \citep{cosmos_ilbert10} shows an agreement within 0.1 dex, indicating that within the uncertainties the AGN contribution is negligible.

\begin{figure}[b]
\centering
\resizebox{0.45\textwidth}{!}{\includegraphics{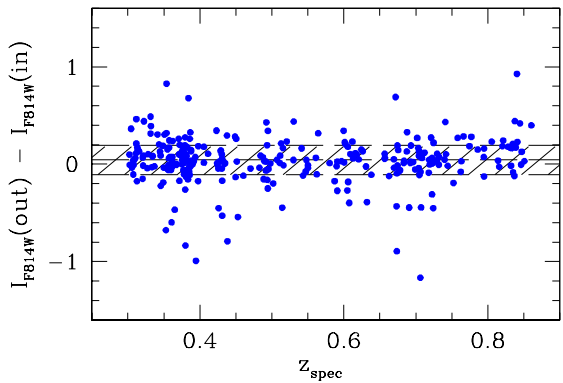}}
\caption{Difference in the observed magnitudes ($I_{F814W}$) of the comparison galaxies before (in) and after (out) the point source addition/subtraction, as a function of redshift.
The $1\sigma$ deviation away from the median is 0.15 mag, indicated by the shaded area centered at 0.01 mag.\label{fig_imp}}
\end{figure}

We require an observable quantity common to both our type-1 and type-2 samples.
For this we define an instrumental, partially k-corrected luminosity based on the photometry from the {\it HST}/ACS imaging with the F814W filter,
\begin{equation}\label{lum}
 L_{F814W}=4\pi\cdot\,d_{L}^2\cdot\,f_{F814W}\cdot(1+z)^{-1}
\end{equation} 
where $f_{F814W}$ is the aperture corrected flux and $d_{L}$ is the luminosity distance.
The $(1+z)^{-1}$ term accounts for the bandpass shifting.

In order to determine a relation between $M_{\ast}/L$ and both redshift and luminosity, and to reduce the covariance between the two variables, we perform a variance-weighted least squares bivariate fit of the form
\begin{equation}\label{m2l}
\mathrm{log}\,M_{\ast}/L_{\mathrm{F814W}}=A\cdot\,z+B\cdot\mathrm{log}(L_{\mathrm{F814W}}/L_{0})+C,
\end{equation}
where $\mathrm{log}(L_0/L_{\odot})=8.2$ corresponds to the minimum value of the luminosity.
Fitting this function to our 199 type-2 AGN host galaxies, using the propagated uncertainties from both $M_{\ast}$ and $L_{F814W}$ to weight each object, results in the coefficients $A=-0.25\pm0.12$, $B=-0.21\pm0.08$, and $C=1.67\pm0.11$.\\

Figure~\ref{fig_m2l} shows $M_{\ast}/L$ for the type-2 AGN sample (filled circles) as a function of redshift (left panel) and luminosity (right panel) and in comparison to $\sim$40,000 inactive galaxies with $I<23$ from the COSMOS catalog \citep{cosmos_ilbert09}.
The latter separate into the usual red sequence and blue cloud with different $M_{\ast}/L$.
The solid lines are the projections of the best-fit plane from Equation~(\ref{m2l}) at the median luminosity and redshift values respectively.
The typical $1\sigma$ errors on the fit are marked by the dashed lines.

\begin{figure*}[ht]
\centering
\resizebox{0.9\textwidth}{!}{\includegraphics{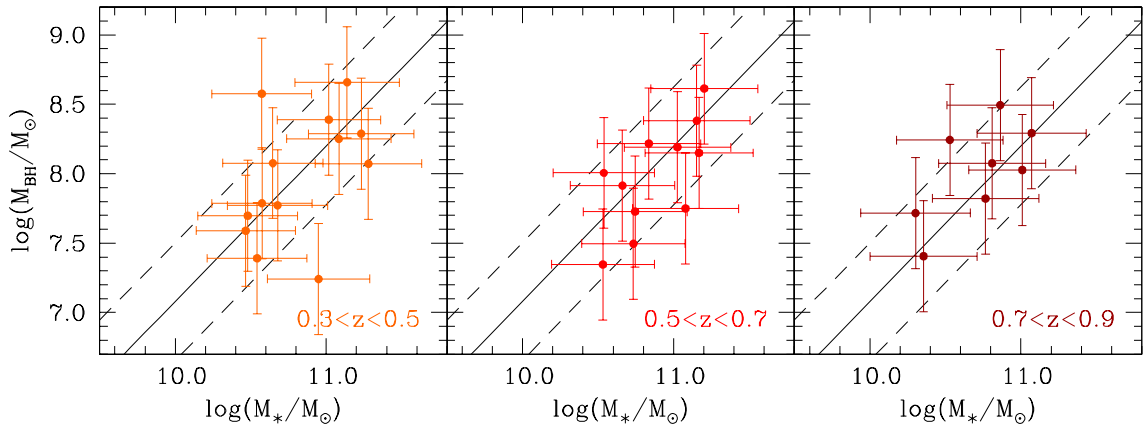}}
\caption{BH mass versus total galaxy stellar mass relation for our sample of 32 type-1 AGNs, shown in three arbitrary redshift bins.
For comparison, the best fit local relation for bulge mass ($\mathrm{log}\,M_{\mathrm{BH}}=1.12\,\mathrm{log}M_{\ast}-4.12$, HR04) is shown.
\label{fig_mm}}
\end{figure*}

\subsection{Type-1 AGN Host Luminosities}

Analyzing type-1 AGN host galaxies remains complicated due to the bright nucleus which can dominate the galaxies' light emission.
Nevertheless, high-resolution {\it HST}/ACS imaging allows us to perform two-dimensional decomposition by modeling the AGN and its host galaxy as a point-spread function (PSF) and a \citet{sersic} profile respectively using GALFIT \citep{galfit02}.
Further multi-component modeling of the host galaxy (bulge+disk) has been shown not to be efficient at our resolution and S/N levels \citep{simmons08}.
The precision of the AGN-host galaxy decomposition strongly depends on the PSF of choice to be supplied to GALFIT.
Benefiting from the COSMOS {\it HST}/ACS area, we can account for instrumental and temporal variations by building specific PSFs for each object by averaging its nearest $\sim$30 stars \citep{jahnke04b}.
For each object, we perform several GALFIT runs with three alternatives for the S\'{e}rsic index $n$: fixed to a $n=1$ exponential profile, to a $n=4$ de Vaucouleurs profile, and also as a free parameter.
The best model is chosen based on a combination of the least $\chi^2$ and a sanity check on the resulting parameters.
We require our galaxy model to have a half-light radius between 2.5 and 100 pixels, i.e., not unphysically large or concentrated; to have a realistic elongation, implying $b/a>0.5$; and for the free $n$ case, not to run outside the range 0.5--8.
After subtraction of the best-model PSF, we are left with the host galaxy emission plus some residuals.

To test whether our image decomposition is over- or undersubtracting point source flux from the host galaxy, we repeat the procedure on a comparison sample of simulated AGN host galaxies \citep[see Section 3.1 of][]{cisternas11a}.
For each type-1 host we select 10 inactive galaxies from the COSMOS catalog matching both in redshift and apparent magnitude.
To each comparison galaxy we add a star as a fake active nucleus, keeping the contrast between host and nucleus of the corresponding type-1 system.
We model and remove the point source as for the original AGNs, and perform photometry on the comparison galaxies before and after the procedure.
Figure~\ref{fig_imp} shows the difference between the initial and recovered magnitudes for the hosts as a function of redshift.
We find only a modest offset of $0.04\pm0.15$ mag, for which we correct our type-1 host galaxy photometry as well as its error budget.
The host galaxy flux will be used to estimate the luminosity as in equation~(\ref{lum}), which is subsequently applied to our derived $M_{\ast}/L$ relation from equation~(\ref{m2l}) to estimate the masses.
The resulting luminosities and stellar masses are presented in Table~\ref{tbl-1}.
The uncertainty in the photometry together with the errors on the fitting coefficients result in a total stellar mass uncertainty of 0.35 dex.

\section{Results and Discussion}

The $M_{\mathrm{BH}}$--$M_{\ast}$ relation for our 32 type-1 AGNs is shown in Figure~\ref{fig_mm} in three redshift intervals.
As a reference, the local relation between BH and bulge mass (solid line, HR04) is shown together with its scatter of 0.3 dex (dashed lines).
The vast majority of our sources (30/32) fall directly within uncertainties in the $z=0$ relation, {\em but with their total instead of bulge stellar mass}.
Our sample presents a median offset perpendicular to the local relation of $\Delta\mathrm{log}(M_{\mathrm{BH}}/M_{\ast})=0.01\pm0.03$, consistent with zero.
In the top panel of Figure~\ref{fig_evo} we show the offset of each object as a function of redshift.
Interestingly, no trend is observed in the offset as a function of increasing redshift.
Nevertheless, to check for traces of redshift evolution in our sample we force a fit to the functional form $\Delta \mathrm{log}(M_{\mathrm{BH}}/M_{\ast})=\delta\,\mathrm{log}(1+z)$ to our data, again using a weighted least squares method.
We find a best fit with $\delta=0.02\pm0.34$ (solid line) consistent with zero evolution within the scatter.

\subsection{Mass Function Bias}

As the observed offset from the local relation is not redshift-dependent, its origin still needs unveiling:
selection effects or pure random scatter?

It has been pointed out in the literature \citep[e.g.,][]{treu07,lauerbias,merloni10,decarli10} that selection effects have to be taken into account when trying to infer the intrinsic scaling relations from the observed data, else a false signal of evolution could be perceived.
While local studies select their samples of inactive galaxies based on galaxy properties ($M_{\ast}$, $L$, $\sigma_{\ast}$), higher redshift samples are selected based on AGN luminosity and hence BH mass, implying that for a given $M_{\mathrm{BH}}$ there is a range of potential stellar masses $M_{\ast}\pm\,dM_{\ast}$ due to an intrinsic cosmic scatter $\sigma_{\mu}$.
If the expected $M_{\ast}$ for a given $M_{\mathrm{BH}}$ happens to be in the steep massive part of the galaxy stellar mass function $\phi(M_{\ast})$ there will be a much higher probability of retrieving a less massive galaxy, which automatically translates into a positive measured offset in $M_{\mathrm{BH}}/M_{\ast}$.

\begin{figure}[ht]
\centering
\resizebox{0.43\textwidth}{!}{\includegraphics{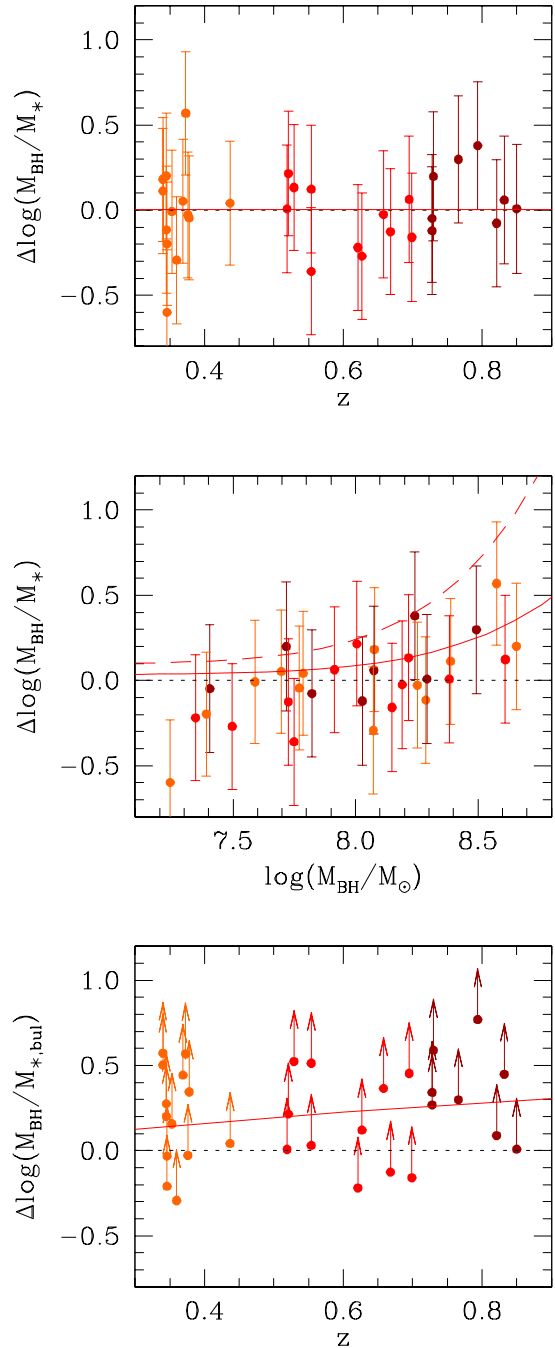}}
\caption{\textbf{Top:} Perpendicular offset from the local relation (dotted line) as a function of redshift.
Fitting a redshift evolution of the form of $\Delta \mathrm{log}(M_{\mathrm{BH}}/M_{\ast})=\delta(1+z)$ yields $\delta=0.02\pm0.34$, shown as the solid line.
\textbf{Middle:} Offset from the local relation as a function of $M_{\mathrm{BH}}$. The expected mass function bias from Equation~(\ref{bias}) is shown for two different cases of intrinsic cosmic scatter: 0.3 dex (solid line), and 0.5 dex (dashed line).
\textbf{Bottom:} As in the top panel, but considering the $bulge$ mass upper limit of the galaxies (see text for our $B/T$ definition). The solid line corresponds to a redshift evolution with $\delta=1.15\pm0.34$.
\label{fig_evo}}
\end{figure}

As in \citet{merloni10}, we quantify the bias based on the result derived by \citet{lauerbias}.
Assuming that the local relation from HR04 holds true within the redshift range probed here, and that $\sigma_{\mu}$ is not too large, the offset as a function of $M_{\mathrm{BH}}$ due to selection effects can be approximated as
\begin{equation}\label{bias}
\Delta\mathrm{log}(M_{\mathrm{BH}}/M_{\ast})\,\approx\,\sigma_{\mu}^2\,\left[\,\frac{d\,\mathrm{ln}\,\phi(M_{\ast})}{d\,\mathrm{log}M_{\ast}}\right]_{M_{\ast}(M_{\mathrm{BH}})}
\end{equation}
where the galaxy mass is given simply by ${\mathrm{log}M_{\ast}=(\mathrm{log}M_{\mathrm{BH}}+4.12)/1.12}$, and $\phi(M_{\ast})$ is the galaxy mass function from the S-COSMOS survey at $z\sim0.5$ \citep{cosmos_ilbert10}.
In the middle panel of Figure~\ref{fig_evo} we show the offset of our sources from the local relation as a function of $M_{\mathrm{BH}}$, as well as the expected offset due to the mass function bias from Equation~(\ref{bias}) for two different values of $\sigma_{\mu}$:
0.3 (solid line) and a more conservative 0.5 (dashed line), which is a good representation of the range of $\sigma_{\mu}$ estimates in the local relation \citep{gultekin09}.

It is clear that even for the likely $\sigma_{\mu}=0.3$ case the mild positive offset at the high-$M_{\mathrm{BH}}$ end can be explained by the expected mass function bias, while the effect is small for the lower-mass half of our sample.
This reaffirms that no signs of evolution are present with respect to total galaxy mass out to $z\sim0.9$.

\subsection{No Significant $M_{\mathrm{BH}}$ and $M_{\ast}$ Growth Down to $z=0$}

Although our objects are in general agreement with the local $M_{\mathrm{BH}}/M_{\ast}$ ratio, they are still observed $3-7$ Gyr before $z=0$.
What constraints can be inferred for their subsequent evolution, both in stellar and BH mass?

From the perspective of the ongoing BH growth, we can estimate the increase in $M_{\mathrm{BH}}$ during the current AGN episode:
(i) Taking an average accretion rate $L/L_{Edd}$ of 0.1 for these sources at $z<1$ \citep{cosmos_trump09b}, and considering a remaining accretion period of half a typical observable AGN lifetime ($\sim$10 Myr) \citep[e.g.,][]{martini04_lifetime,porciani04}, the increase of log$M_{\mathrm{BH}}$ will be $\sim$0.001 dex.
(ii) A definite upper limit on the subsequent BH growth can be given by considering the AGN lifetime derived from the number density of XMM-COSMOS AGNs by \citet{gilli09}.
Out to $z\sim1$ they compute an AGN duty cycle of 0.1, corresponding to an AGN lifetime of $\sim$1 Gyr.
Using the same $L/L_{Edd}$ as before, we estimate that our sources will {\em at most} grow 0.09 dex in BH mass due to accretion.
This implies that these sources will hardly move in the upward direction of the $M_{\mathrm{BH}}$--$M_{\ast}$ plane.

Concurrently, we can give upper limits to the stellar mass increase due to star formation in these galaxies.
Based solely on the average specific star formation rates at this mass range ($\mathrm{log}(M_{\ast}/M_{\odot})=10.2-11.2$; \citealt{karim11}), from their respective redshift down to $z=0$, in the most extreme case the increase will be a factor of 1.8 (or $0.25$ dex).
This ignores potential quenching mechanisms and is sufficient to show that we should not expect a significant change in stellar mass.

Regarding merger activity, the major merger rate (mass ratio $>$1:3) since $z\sim0.5$ is low ($\sim$0.05 Mergers Gyr$^{-1}$; \citealt{hopkins10_mergers}) and should
(i) not have a significant impact on the total galaxy mass of the ensemble, and
(ii) major mergers will add similar mass fractions to both galaxy and BH.

Nevertheless, the true relevance of the overall, mainly minor merging activity will be in the subsequent bulge growth of these galaxies:
The fact that the {\em total} stellar mass of our galaxies is already consistent with the local relation for galactic bulges at $z\sim0.7$ means that, for them to obey the local relation by $z=0$, their current total mass should end up redistributed as bulge mass.
Secular processes such as minor interactions and internal instabilities are the prevailing bulge-building mechanisms \citep{kormendy&kennicutt04,weinzirl09,parry09}, and are coincident with the likely mechanisms that must dominate the BH fueling at $z<1$ \citep{cisternas11a}.
This has the implication that, over the last 7 Gyr, no extreme mass growth is required to produce the local scaling relations.
Instead a rather passive, non-violent secular evolution will drive the redistribution of mass into the bulge components.

With this in mind, our results do not necessarily contradict previously reported evolution with respect to {\em bulge} properties in the redshift windows $z=0.36,0.57$ \citep[e.g.,][]{treu07,woo08,bennert10}.
Based on our GALFIT modeling, we can make some basic assumptions to estimate the bulge-to-total mass ratio ($B/T$) of our galaxies:
assuming a $B/T$ upper limit of 0.3 and 0.6 for the galaxies with $n<2$ and $2<n<3$ respectively, we can give a rough estimate of $M_{\ast,bul}$ at least for illustrative purposes.
For the $n>3$ case, we take an upper limit of $B/T$=1, as it corresponds to a bulge-dominated galaxy.
These estimates are consistent with the recovered $n$ values for simulated $B/T$ distributions by \citet{simmons08}.
According to our definition of $M_{\ast,bul}$, we show its offset as a function of redshift in the bottom panel of Figure~\ref{fig_evo}.
Fitting the same functional form as before yields $\delta=1.15\pm0.34$, an indication that our data permits an evolution with respect to $bulge$ mass, and in broad agreement with \citet{treu07}.

\section{Conclusions}

A representative AGN host galaxy $M_{\ast}/L$ was used to estimate stellar masses for 32 type-1 AGN hosts at $0.3<z<0.9$.
Combined with already available virial $M_{\mathrm{BH}}$ estimates, we studied the $M_{\mathrm{BH}}-M_{\ast}$ relation out to 7 Gyr lookback time,
and extended recent studies probing the $z>1$ regime \citep{jahnke09,merloni10}.
In summary:

\begin{itemize}
\item[1.] $M_{\ast}/L$ for intermediately luminous AGNs at a given redshift and luminosity has scatter of only $\sim$0.25 dex.

\item[2.] Within a 0.03 dex uncertainty in the mean, the {\em total} mass of our sources is consistent with zero offset from the $z=0$ relation for galactic {\em bulges}.

\item[3.] No increase of the offset was found with redshift. Nevertheless, a forced fit to the functional form $\delta\,\mathrm{log}(1+z)$ yields $\delta=0.02\pm0.34$, confirming non-evolution.

\item[4.] We found that a positive offset exists in the observed value at high $M_{\mathrm{BH}}$. When including the mass function bias upon inference of the intrinsic relation for our sample, we found that this selection effect accounts for the increasing offset with $M_{\mathrm{BH}}$.

\item[5.] 
The fact that the majority of these galaxies are disk-dominated, together with the lack of evolution with respect to total $M_{\ast}$, implies that all mass to be found in the bulge at $z=0$ is already present in the galaxy at the observed redshift, and the only process required is a redistribution of stellar mass from disk to bulge driven by secular evolution, in agreement with the dominating AGN triggering mechanisms predicted by \citet{cisternas11a}.

\item[6.] Our result allows an evolution of the bulge mass scaling relations: a simple conservative assessment of $B/T$ based on our GALFIT modeling yields $\delta=1.13\pm0.34$.

\end{itemize}

We presented a simple and solid technique to estimate stellar masses of type-1 AGN host galaxies when no high-resolution multiwavelength coverage is present.
Our work explicitly confirmed the expected non-evolution of the $M_{\mathrm{BH}}-M_{\ast}$ ratio out to $z$=0.9.
The low levels of star formation, merger activity, and BH growth do not allow for any extreme evolution at these redshifts and secular processes must dominate any changes in the mass distributions and structures of galaxies.


\acknowledgments
M.C. thanks Anna Gallazzi for helpful, thorough assessments, and Brooke Simmons for practical comments.
M.C., K.J., and K.I. are supported through the Emmy Noether Programme of the German Science Foundation (DFG).

{\it Facilities:} \facility{{\it HST} (ACS)},
\facility{Magellan (IMACS)}, \facility{{\it XMM-Newton}}.





\newcommand{\noopsort}[1]{}

\end{document}